\documentclass[aps,epsfig,twocolumn]{revtex4}

\usepackage{graphicx}
\usepackage{dcolumn}
\usepackage{bm}

\newcommand{\bq}{\begin{equation}}
\newcommand{\eq}{\end{equation}}
\newcommand{\bqa}{\begin{eqnarray}}
\newcommand{\eqa}{\end{eqnarray}}
\newcommand{\nn}{\nonumber \\}

\begin{document}
\draft
\title{
Emergence of supersymmetry at a critical point of a lattice model
}

\author{
Sung-Sik Lee
}
\address{
Kavli Institute for Theoretical Physics, University of California,\\
Santa Barbara, California 93106, U.S.A.\\
}
\date{\today}

\date{\today}

\begin{abstract}

Supersymmetry is a symmetry between a boson and a fermion.
Although there is no apparent supersymmetry in nature,
its mathematical consistency and appealing property
have led many people to believe that
supersymmetry may exist in nature in the form of 
a spontaneously broken symmetry.
In this paper, we explore an alternative possibility
by which supersymmetry is realized in nature, that is, 
supersymmetry dynamically emerges in the low energy limit 
of a non-supersymmetric condensed matter system.
We propose a 2+1D lattice model 
which exhibits an emergent space-time supersymmetry 
at a quantum critical point.
It is shown that there is only one relevant perturbation 
at the supersymmetric critical point 
in the $\epsilon$-expansion and 
the critical theory is the two copies of the 
Wess-Zumino theory with four supercharges. 
Exact critical exponents are predicted.
\end{abstract}
\maketitle

\section{Introduction}
Poincare invariance is the underlying space-time 
symmetry of relativistic quantum field theories.
There are only two mathematically consistent ways of extending 
the symmetry in nontrivial ways due to a no-go theorem\cite{CM}. 
One is a conformal symmetry and the other, a supersymmetry.
A conformal symmetry combines the Poincare invariance
with scale invariance.
It is realized in the long-distance limit of a massless theory 
at which all finite length scales are scaled out. 
A supersymmetry is a symmetry between a boson and a fermion.
Generators of supersymmetry $Q_\alpha$, which are called supercharges, are spinors and they satisfy the commutation relations,
\bqa
\{ Q_\alpha, Q_\beta \} & = & 2 P_\mu \Gamma^\mu_{\alpha \beta}, \nn
  \left[ Q_\alpha, P_\mu \right] & = & 0,
 \label{Q}
\eqa
where $\alpha$, $\beta$ are spinor indices, 
$\Gamma^\mu_{\alpha \beta}$ are constants, 
and $P_\mu$ is the energy-momentum operator.
Because supercharges are fermionic operators,
a boson is transformed into a fermion (and vice versa) 
under supersymmetry transformations.
Therefore the number of bosonic modes is equal
to the number of fermionic modes in supersymmetric theories.
The second commutation relation in Eq. (\ref{Q}) implies 
$[P_\mu P^\mu, Q_\alpha]=0$ and masses of supersymmetric partners
are identical.
Since bosons and fermions contribute to quantum effective actions
with quantum corrections of the opposite signs
and they have same masses, 
the effects of quantum fluctuations of bosons and fermions are canceled 
with each other in supersymmetric theories.
Because of this, quantum corrections are highly constrained by kinematics.
If there are enough supersymmetries, 
there is no quantum correction at all (non-renormalization) 
for some quantities.
Due to the non-renormalization property,
supersymmetry has been proposed as a  
way of stabilizing the hierarchy of vastly different 
mass scales present in the standard model. 
It may also play an important role in the unification of the gauge interactions.
On the other hand, many supersymmetric theories 
have been studied as toy models where
a supersymmetry enables one to understand
strong coupling physics rigorously\cite{SW,SEIBERG}.

Despite the unique mathematical consistency and 
beautiful properties of supersymmetry, 
nature does not exhibit supersymmetry at low energy scales. 
If nature is supersymmetric, it should be spontaneously broken.
If that is the case, supersymmetry will become manifest
at a high energy scale.
An alternative way of finding supersymmetry in nature
may be to go to a low energy in condensed matter systems, 
relying on the principle of emergence.

A new symmetry can emerge in the low-energy limit 
although a microscopic model does not respect the symmetry.
For example, a low-energy effective theory 
can have the full Poincare invariance 
although the underlying lattice explicitly breaks
rotational symmetry in a condensed matter system.
Even a gauge symmetry\cite{FOERSTER} and a 
general covariance\cite{GU,LEE} can be emergent.
The emergence of a new symmetry can be a characteristic 
of a new state of many-body systems\cite{WEN}.
Searching for new states of matter in condensed matter systems
is becoming an important research avenue as 
new materials which can not be understood in conventional theories 
are synthesized\cite{SHIMIZU1,HELTON} and
highly controllable correlated many-body systems
can be fabricated in cold atom systems\cite{ANDERSON,DAVIS}.
Therefore it would be of interest to find a condensed matter system
which shows an emergent supersymmetry.

It has been suggested that supersymmetry can 
emerge in the low-energy limit of a non-supersymmetric theory\cite{CURCI, GOH,SCOTT}.
In 1+1D, supersymmetry emerges at the tricritical point of the dilute Ising model\cite{FRIEDAN}.
Emergent supersymmetries play important roles in realizing lattice versions of supersymmetric field theories in various dimensions\cite{FEO}.
For example, the ${\cal N}=1$ 3+1D super Yang-Mills theory can emerge without an underlying supersymmetry although the notion of the emergent supersymmetry is rather obscure in this case due to the opening of a mass gap caused by confinement.
Supersymmetric field theories can be also realized 
in lattices by fine tunings of bare parameters\cite{ELLIOTT} 
or by a dynamical mechanism where some supersymmetries 
which are already present in lattices guarantee 
the emergence of continuum supersymmetries without fine tuning\cite{REVIEW}.
In this paper, we construct a 2+1D lattice model where supersymmetry may
dynamically emerge at a critical point without any lattice supersymmetry.

It is noted that non-relativistic supersymmetries have been considered in condensed matter systems\cite{FORSTER,FENDLEY}.
In non-relativistic systems, supercharges are scalars (not spinors) and the anti-commutator of supercharges generates only energy (not momentum), that is, $\{ Q, Q^\dagger \} = H$, where $H$ is a Hamiltonian.
In such systems, supersymmetries play the roles as in 0+1D quantum mechanical systems\cite{susyQM}.
The present work concerns an emergence of a full space-time supersymmetry in a 2+1D relativistic system where the relativity is also emergent out of a non-relativistic microscopic system.
In this case, the algebra of supercharges generate the translations in both time and space through the energy-momentum operator as in Eq. (\ref{Q}).

\section{Microscopic model and low-energy effective theory}
\begin{figure}[h!]
        \includegraphics[height=11cm,width=9cm]{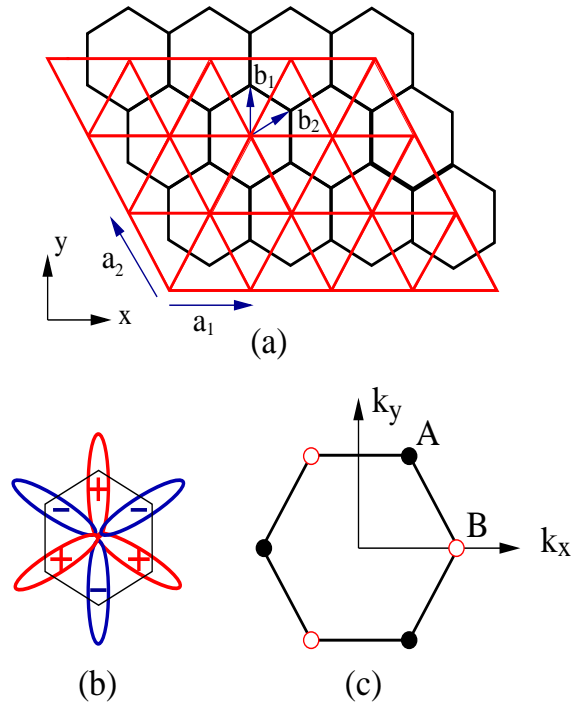}
\caption{
(a) The lattice structure in the real space. 
Fermions are defined
on the honeycomb lattice and 
the bosons, on the dual triangular lattice. 
${\bf a}_1$, ${\bf a}_2$ are the lattice vectors with length $a$, and
${\bf b}_1$, ${\bf b}_2$, two independent vectors
which connect a site on the triangular lattice
to the nearest neighbor sites on the honeycomb lattice.
(b) The phases of a fermion pair in the real space.
(c) The first Brillouin zone in the momentum space.
$A$ and $B$ indicate two inequivalent points with momenta 
${\bf k}_A = \frac{2 \pi}{a} ( \frac{1}{3}, \frac{1}{\sqrt{3}} )$ 
and ${\bf k}_B = \frac{2 \pi}{a} ( \frac{2}{3}, 0 )$ 
where the low energy modes are located.
$\psi_1$, $\phi_2$ are located at ${\bf k}_A$ and
$\psi_2$, $\phi_1$, at ${\bf k}_B$.
}
\label{fig:1}
\end{figure}

The microscopic system is a mixture of fermions and bosons.
The Hamiltonian is composed of three parts, 
\bqa
H & = & H_f + H_b + H_{fb},  
\label{model}
\eqa
where
\bqa
H_f & = & -t_f \sum_{<i,j>} ( f_i^\dagger f_j + h.c. ), \nn
H_b & = & t_b \sum_{<I,J>} ( e^{ i ( \theta_I - \theta_J )} + h.c. ) + \frac{U}{2} \sum_I n_I^2, \nn
H_{fb} & = & h_0 \sum_I e^{i \theta_I} (  
f_{I+ {\bf b}_1} f_{I-{\bf b}_1} +     f_{I-{\bf b}_2} f_{I+{\bf b}_2} +  \nn
&& + f_{I-{\bf b}_1+{\bf b}_2} f_{I+{\bf b}_1-{\bf b}_2}     ) + h.c.. 
\eqa
Here $H_f$ describes spinless fermions with nearest neighbor hopping on the honeycomb lattice;
$H_b$ describes bosons with nearest neighbor hopping and an on-site repulsion on the triangular lattice 
which is dual to the honeycomb lattice; and 
$H_{fb}$ couples the fermions and bosons.
The lattice structure is shown in Fig. \ref{fig:1} (a).
$f_i$ is the fermion annihilation operator and
$e^{-i \theta_I}$, the lowering operator of $n_I$
which is conjugate to the angular variable $\theta_I$.
$i,j$ and $I,J$ are site indices for the honeycomb and triangular lattices, respectively.
$t_f, t_b  > 0$ are the hopping energies for the fermions and bosons, respectively 
and $U$ is the on-site boson repulsion energy.
Note that the boson hopping is frustrated.
This will play a crucial role for the emergent supersymmetry as will be shown later.
${\bf b}_1$ and ${\bf b}_2$ are vectors which connect 
a site on the triangular lattice to the neighboring honeycomb lattice sites 
as is shown in Fig. \ref{fig:1} (a).
$h_0$ is the pairing interaction strength associated with the process
where two fermions in the f-wave channel around a hexagon 
are paired and become a boson at the center of the hexagon, and vice versa.
In this sense, the boson can be regarded as a Cooper pair 
made of two spinless fermions in the f-wave wavefunction
as is shown in Fig. \ref{fig:1} (b).
This model has a global U(1) symmetry
under which the fields transform as
$f_i \rightarrow f_i e^{i \varphi}$ and 
$e^{-i \theta_I}  \rightarrow e^{-i \theta_I} e^{i 2 \varphi}$.

First, we identify low-energy modes of the fermions and bosons in the absence of the coupling $H_{fb}$.
At zero chemical potential, the fermions are half-filled and
their energy spectrum is given by
$e^f_{\bf k} = \pm t_f 
\sqrt{  
( 1 + \cos  k_1   + \cos k_2 )^2
+ ( \sin k_1 - \sin k_2  )^2
}$
where $k_1 = a k_x$, $k_2 = a \frac{ -k_x + \sqrt{3} k_y}{2} $
with $a$, the lattice spacing.
There exist two Fermi points
at ${\bf k}_A = \frac{2 \pi}{a} ( \frac{1}{3}, \frac{1}{\sqrt{3}} )$ and
 ${\bf k}_B = \frac{2 \pi}{a} ( \frac{2}{3}, 0 )$
as shown in Fig. \ref{fig:1} (c).
Since the energy dispersion is linear near the Fermi points,
the low-energy excitations are described by two Dirac fermions,
\bqa
{\cal L}_f & = & i \sum_{n=1}^2 
{\overline \psi_n} 
\left(
\gamma_0 \partial_\tau +  c_f \sum_{i=1}^2  \gamma_i \partial_i  
\right) \psi_n.
\label{lf}
\eqa
Here $\psi_1$ ($\psi_2$) denotes the two-component complex fermion 
at momentum ${\bf k}_A$ (${\bf k}_B$). 
$\partial_\mu = (\partial_\tau, \partial_x, \partial_y)$ are the derivatives
in imaginary time and the spatial directions.
$\gamma_0 \equiv \sigma_3$, 
$\gamma_1 \equiv \sigma_1$ 
and $\gamma_2 \equiv \sigma_2$ 
with $\sigma_\mu$, the Pauli matrices.
${\overline \psi_n} \equiv -i \psi_n^\dagger \gamma_0$ and 
$c_f \sim t_f a $ is the fermi velocity.

To obtain a low-energy theory for the bosons, 
we introduce a soft boson field $\Phi_I = |\Phi_I | e^{- i \theta_I}$
and the potential $V(\Phi) = u_2 |\Phi|^2 + u_4 |\Phi|^4$ 
which gives a finite amplitude to the soft boson field.
In energy-momentum space, the boson action becomes
\bqa
{\cal S}_b & = & \int dk \left( \frac{1}{2U} k_0^2 + e^b_{\bf k} + u_2 \right) |\Phi_k|^2 \nn
& + & u_4 \int dk_1 dk_2 dq  ~  \Phi_{k_2-q}^* \Phi_{k_1+q}^* \Phi_{k_1} \Phi_{k_2},
\eqa
where $\int dk \equiv \int \frac{dk_0 dk_x dk_y}{(2\pi)^3}$ is the energy-momentum integration
and
$e^b_{\bf k} = 2 t_b \left[ \cos k_1 
+ \cos k_2    
+  \cos (k_1 + k_2 ) 
\right]$. 
Since the boson hopping has the wrong sign, 
${\bf k}=(0,0)$ is not the minimum of $e^b_{\bf k}$;
rather, two minima occur at ${\bf k}_A$ and ${\bf k}_B$ 
where the nodal points of the fermions are located.
Therefore we have two low-energy boson modes.
We introduce $\phi_1$ and $\phi_2$
to represent the low-energy modes near the ${\bf k}_B$ and 
${\bf k}_A$ points, respectively.
Note that the $\phi_1$ ($\phi_2$) boson carries 
the same momentum as the $\psi_2$ ($\psi_1$) fermion.
With this convention, we will see that
only those bosons and fermions which carry
the same index ($n=1,2$) interact with each other 
at low energies.
Expanding $e^b_{\bf k}$ near the two minima, we obtain
the effective Lagrangian for the low-energy bosons,
\bqa
{\cal L}_b & = & \sum_{n=1}^2 
\left[  
 |\partial_\tau \phi_n|^2 + c_b^2 \sum_{i=1}^2 |\partial_i \phi_n|^2
+ m^2 |\phi_n|^2 \right] \nn
&&  + \lambda_1 \kappa \sum_{n=1}^2  |\phi_n|^4 
    + \lambda_2  \kappa |\phi_1|^2 |\phi_2|^2,
\label{lb}
\eqa
where $c_b \sim \sqrt{ t_b U} a$ is the boson velocity 
which is in general different from the
fermion velocity $c_f$.
Although both the fermions and bosons have the `relativistic' energy spectrums,
there is no Lorentz symmetry if the velocities are different.
The Lorentz symmetry requires the velocities of all massless particles 
to be identical.
As will be shown later, the Lorentz symmetry will emerge 
in the low energy limit through quantum corrections.
$m$ is the boson mass and
the coupling constants $\lambda_1$, $\lambda_2$ 
are made dimensionless
by introducing a mass scale $\kappa$.
Note that momentum conservation does not allow an
interaction such as $\phi_2^* \phi_2^* \phi_1 \phi_1$.

We can obtain the interaction between the low-energy fermions and bosons 
by rewriting $H_{fb}$ in energy-momentum space and
keeping only the low-energy modes.
The resulting interaction Lagrangian is 
\bqa
{\cal L}_{fb} & = & h \kappa^{1/2}
 \sum_{n=1}^2 \left( \phi_n^* \psi_n^T \varepsilon \psi_n +   c.c.  \right),
\label{lfb}
\eqa
where $\varepsilon$ is the $2 \times 2$ antisymmetric matrix
with $\varepsilon_{12} = - \varepsilon_{21} = 1$.
Terms like 
$\phi_2^* \psi_1^T \varepsilon \psi_1$ 
or
$\phi_2^* \psi_1^T \varepsilon \psi_2$ 
are not allowed 
because they do not satisfy momentum conservation.

\section{Renormalization group analysis}
Now we perform a one-loop 
renormalization group (RG) analysis in $4-\epsilon$ dimensions
for the  low-energy effective theory given by
\bqa
{\cal L} & = & {\cal L}_f +  {\cal L}_b + {\cal L}_{fb}. 
\eqa
We use the dimensional regularization scheme where
the number of fermion components and the traces of gamma matrices
are fixed\cite{TOWNSEND}.
Maintaining the same number of fermionic and bosonic modes in $4-\epsilon$ dimension
is important because supersymmetry requires that the number of modes is the same for the bosons and fermions. 
If there is a gauge symmetry, more sophisticated regularization scheme is necessary to
preserve both gauge symmetry and supersymmetry
because the number of components of gauge boson should depend on the dimension of space-time\cite{STOCK}.
Since there is no gauge symmetry in the present model,
the simple dimensional regularization scheme can maintain
supersymmetry\cite{JJ}.
Of course, the present model has no supersymmetry.
The point is that it is convenient to use a regularization scheme which can maintain supersymmetry in probing an emergent supersymmetry in the low energy limit.

\begin{figure}[h!]
        \includegraphics[height=11cm,width=8.5cm]{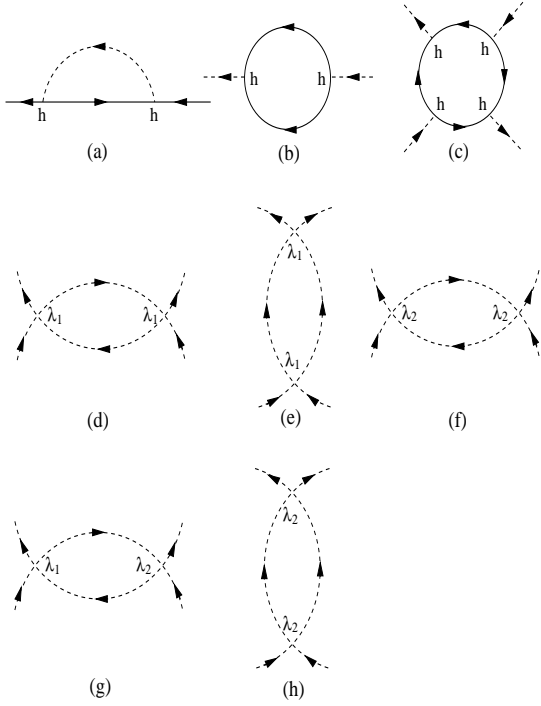}
\caption{
One-loop diagrams.
(a) and (b) are the self-energy corrections of fermion
and boson respectively. 
(c), (d), (e) and (f) contribute to the vertex correction 
of $\lambda_1$ and
(f), (g) and (h), to the vertex correction of $\lambda_2$.
}
\label{fig:2}
\end{figure}

In the $\epsilon$-expansion, the above Lagrangian
contains all the relevant and marginal terms.
A four fermion interaction has the scaling dimension
$D = 6 - 2 \epsilon + O(\epsilon^2)$ and can be ignored for a small $\epsilon$.
In the following, we do not consider the four fermion interaction.
However, in principle, the four fermion interaction can become important in $2+1D$ 
due to a strong interaction, in which case
one needs to tune a microscopic four fermion interaction term to reach 
the fixed points we will discuss in the following.
The boson mass is always a relevant perturbation and 
we tune it to zero in order to examine the RG 
flow of the other couplings in the massless subspace.
At the one loop-level, there are $8$ diagrams which are shown in Fig. \ref{fig:2}.
Each diagram contributes to the quantum effective action as follows,
\bqa
\delta {\cal L}^{(a)} & = &
\frac{ 4h^2}{(4\pi)^2 c_f^2 \epsilon} \sum_{n=1}^2  i \bar \psi_n \left( f_0 \gamma_0 \partial_\tau + c_f f_1 \sum_{i=1}^2  \gamma_i \partial_i  \right) \psi_n, \nn
\delta {\cal L}^{(b)} & = &
\frac{ 4h^2}{(4\pi)^2 c_f^2 \epsilon} \sum_{n=1}^2   \left(  | \partial_\tau \phi_n|^2  + c_f^2  
\sum_{i=1}^2 | \partial_i \phi_n|^2  \right), \nn
\delta {\cal L}^{(c)} & = &
\frac{ 16 h^4}{(4\pi)^2 c_f^2 \epsilon} \sum_{n=1}^2  
| \phi_n|^4, \nn
\delta {\cal L}^{(d)} & = & 
- \frac{ 16 \lambda_1^2}{(4\pi)^2 c_b^2 \epsilon} \sum_{n=1}^2
| \phi_n|^4, \nn
\delta {\cal L}^{(e)} & = & 
- \frac{ 4 \lambda_1^2}{(4\pi)^2 c_b^2 \epsilon} \sum_{n=1}^2
| \phi_n|^4, \nn
\delta {\cal L}^{(f)} & = & 
- \frac{  \lambda_2^2}{(4\pi)^2 c_b^2 \epsilon} \sum_{n=1}^2 
| \phi_n|^4   - \frac{  2 \lambda_2^2}{(4\pi)^2 c_b^2 \epsilon} 
| \phi_1|^2 | \phi_2|^2, \nn
\delta {\cal L}^{(g)} & = & 
- \frac{ 16 \lambda_1 \lambda_2}{(4\pi)^2 c_b^2 \epsilon}  
| \phi_1|^2 | \phi_2|^2, \nn
\delta {\cal L}^{(h)} & = & 
- \frac{ 2 \lambda_2^2}{(4\pi)^2 c_b^2 \epsilon}
| \phi_1|^2 | \phi_2|^2, \nn
\eqa
where
$f_0 = \frac{4}{\alpha(\alpha+1)^2}$ and $f_1 = \frac{4 (2\alpha+1)}{3\alpha(\alpha+1)^2}$ with $\alpha = c_b/c_f$.
From the renormalized quantum effective action, the beta functions are obtained to be
\bqa
\frac{d h}{d l} & = & \frac{\epsilon}{2} h - \frac{ 1}{ ( 4 \pi c_f )^2 } 
\left( 2 + \frac{ 16 c_f^3}{ c_b ( c_f + c_b )^2} \right) h^3,  \nn
\frac{d \lambda_1}{d l} & = & \epsilon \lambda_1 - \frac{ 1 }{ ( 4 \pi )^2 } 
\left( 
\frac{ 20 \lambda_1^2 + \lambda_2^2 }{ c_b^2} 
+\frac{ 8 h^2 \lambda_1 }{ c_f^2} 
-\frac{ 16 h^4 }{ c_f^2} 
\right), \nn
\frac{d \lambda_2}{d l} & = & \epsilon \lambda_2 - \frac{ 1 }{ ( 4 \pi )^2 } 
\left( 
\frac{ 4 \lambda_2^2 + 16 \lambda_1 \lambda_2 }{ c_b^2} 
+\frac{ 8 h^2 \lambda_2 }{ c_f^2} 
\right), \nn
\frac{d c_f}{d l} & = & \frac{ 32 h^2 c_f ( c_b - c_f ) }{ 3 ( 4 \pi )^2 c_b ( c_b + c_f )^2}, \nn
\frac{d c_b}{d l} & = & -\frac{ 2 h^2 c_b ( c_b^2 - c_f^2 ) }{  ( 4 \pi c_b c_f )^2 },
\label{beta}
\eqa
where the scaling parameter $l$ increases in the infrared.

\begin{figure}[h!]
        \includegraphics[height=4cm,width=5cm]{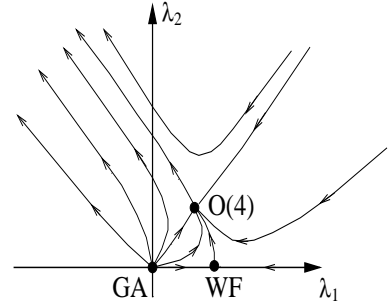}
\caption{
The schematic  RG flow of the bosonic couplings in the subspace 
of $m=h=0$.
}
\label{fig:3}
\end{figure}

There are two solutions for $\beta_h=0$.
One is the unstable solution with $h = 0$
and the other, the stable one with 
a finite $h$.
At $h=0$, the bosons and fermions are decoupled.
The fermion system consists of non-interacting Dirac fermions.
The RG flow of the boson couplings in the subspace of $m=h=0$ is shown in Fig. \ref{fig:3}. 
In the subspace of $m=h=0$, there are three fixed points, that is, 
the Gaussian (GA) fixed point with $( h^*, \lambda_1^*, \lambda_2^* ) = (0,0,0)$,
the Wilson-Fisher (WF) fixed point with 
$( h^*, \lambda_1^*, \lambda_2^* ) = (0,\frac{( 4 \pi c_b)^2 \epsilon}{20},0)$, 
and the O(4) fixed point with
$( h^*, \lambda_1^*, \lambda_2^* ) = 
(0,\frac{( 4 \pi c_b)^2 \epsilon}{24},\frac{( 4 \pi c_b)^2 \epsilon}{12})$.
Because the linear term of the beta function at the O(4) fixed point accidentally  vanishes along a direction, at higher order in $\epsilon$ there occurs a stable fourth fixed point between the WF fixed point and the O(4) fixed point\cite{KAWAMURA}.
However, the fixed points in the $m=h=0$ plane are all unstable because the pairing interaction $h$ is relevant.

\begin{figure}[h!]
        \includegraphics[height=10cm,width=7cm]{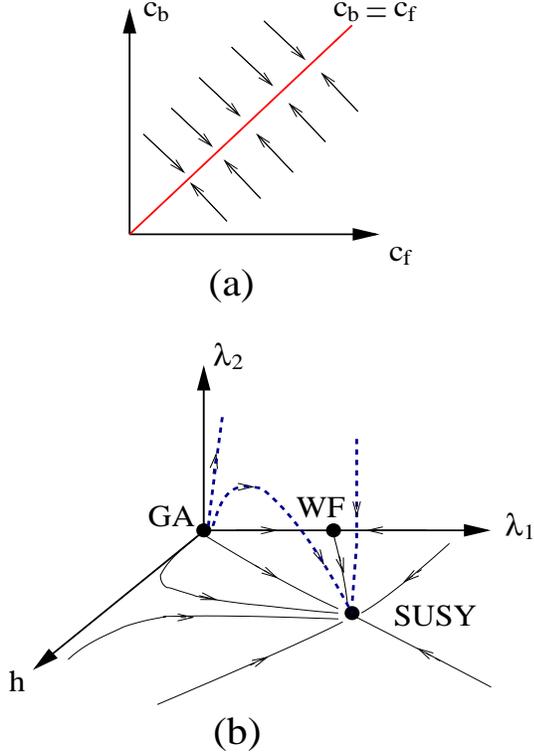}
\caption{
The schematic RG flows of 
(a) the velocities with $h \neq 0$ and
(b) $\lambda_1$, $\lambda_2$ and $h$ in the subspace of $m=0$.
In (b), the solid lines represent the flow in the plane of $(h,\lambda_1)$
and the dashed lines, the flow outside the plane.
}
\label{fig:4}
\end{figure}

Once we turn on the pairing interaction, 
$h$ flows to a finite value with
$h^2 =  \frac{ ( 4 \pi c_f)^2 \epsilon }{ 2 } \left( 2 + \frac{ 16 c_f^3}{ c_b ( c_f + c_b )^2} \right)^{-1}$ and the boson and fermion velocities begin to
flow as can be seen from the last two equations in Eq. (\ref{beta}).
Because the pairing interaction mixes the velocities of the boson and fermion, the difference of the velocities exponentially flows to zero in the low-energy limit 
as is shown in Fig. \ref{fig:3} (a).
The line of $c_b = c_f$ is critical and the value of the velocity in the infrared limit is a non-universal value.
This implies that the bosons and fermions have the same energy dispersion and Lorentz symmetry emerges at low energies due to quantum fluctuations.
Now we consider the flow of $h$, $\lambda_1$ and $\lambda_2$ with a fixed $c_b = c_f = c$.
The RG flow is displayed in Fig. \ref{fig:3} (b).
In the following we will use units where $c = 1$.
With a nonzero $h$, the system flows to a stable fixed point,
\bq
( h^*, \lambda_1^*, \lambda_2^* ) = (\sqrt{\frac{(4 \pi)^2 \epsilon}{12}}, \frac{(4 \pi)^2 \epsilon}{12} ,0).
\eq
The nonzero $h$ is a consequence of strong pairing fluctuations at the critical point.
This is crucial in obtaining Lorentz symmetry and supersymmetry as will be discussed in the followings.
At the fixed point, $\lambda_2$ vanishes and 
there is no coupling between the two sets of low-energy modes, 
$(\phi_n, \psi_n)$ with $n=1,2$.
Physically, this implies that the bose condensates which carry the different momenta ${\bf k}_A$ and ${\bf k}_B$ develop independently in the condensed phase.
At the critical point, $\lambda_1 = h^2$ and 
the theory becomes invariant under 
the transformation,
\bqa
\delta_{\xi_n} \phi_n  & = &  - {\overline \psi_n} \xi_n, \nn
\delta_{\xi_n} \phi^*_n  & = &  {\overline \xi_n} \psi_n, \nn
\delta_{\xi_n} \psi_n & = & i \slash \hspace{-0.2cm} \partial  \phi_n^* \xi_n - \frac{h}{2} \phi_n^2 {\cal \varepsilon} {\overline \xi_n}^T, \nn  
\delta_{\xi_n} {\overline \psi_n} & = & i {\overline \xi_n}  \slash \hspace{-0.2cm} \partial \phi_n - \frac{h}{2} \phi_n^{*2} \xi^T_n {\cal \varepsilon},
\label{susy}
\eqa
where \mbox{$\slash \hspace{-0.2cm} \partial = \gamma_\mu \partial_\mu$} and
$\xi_n$ is a two-component spinor of Grassmann variables which parameterizes the transformation. 
This is a supersymmetry because the bosons and fermions are mixed under the transformation.
Since the two sets of modes $(\phi_n, \psi_n)$ are decoupled at the critical point, 
the supersymmetry transformations are independent for $n=1$ and $2$.
That is why the spinor $\xi_n$ has the index $n$.
Here $\delta_{\xi_n} {\overline \psi_n}  \neq -i ( \delta_{\xi_n} \psi_n )^\dagger \gamma_0$ because we are using the imaginary time formalism. 
The supersymmetry leads to conserved supercurrents,
\bqa
J_\mu^n & = &  \slash\hspace{-0.2cm} \partial \phi_n   \gamma_\mu \psi_n + i \frac{h}{2} \phi_n^2 \gamma_\mu \epsilon {\overline \psi^T_n}, \nn
{\overline J_\mu^n} & = &   {\overline \psi_n} \gamma_\mu \slash\hspace{-0.2cm} \partial \phi_n^*  + i \frac{h}{2} \phi_n^{*2} \psi^T_n \epsilon \gamma_\mu. 
\eqa
Here the spinor indices in $J_\mu^n$ and ${\overline J_\mu^n}$ are suppressed.
For each $n$, there are four independent supercharges, 
$Q_\alpha^n = \int dx^2 J_{0 \alpha}^n$ and 
${\overline Q_\alpha^n}= \int dx^2 {\overline J_{0\alpha}^n}$ in 2+1D.
The supercharges satisfy the commutation relations like Eq. (\ref{Q}).
This corresponds to ${\cal N}=2$ supersymmetry in each sector of $n$, that is,
there are twice as much supercharges as the minimum number of supercharges in 2+1D.
The resulting super-Poincare invariance is emergent  
because the microscopic model has
neither Poincare symmetry nor supersymmetry.
Each set made of one complex boson and one two-component fermion forms a chiral multiplet of the supersymmetry.
This critical theory is the ${\cal N}=2$ Wess-Zumino theory\cite{WZ} with two copies of chiral multiplets\cite{C2}.
The critical exponents calculated in the one-loop level
\bq 
\eta_\phi = \eta_\psi = \epsilon/3
\label{SD}
\eq
match those of the Wess-Zumino theory in the leading order of $\epsilon$.
In the supersymmetric theory, the one-loop result is exact as will be explained later.
It is of note that the critical exponent is independent of regularization scheme although
the values of the couplings depend on regularization scheme.

\begin{figure}[h!]
        \includegraphics[height=2.5cm,width=8cm]{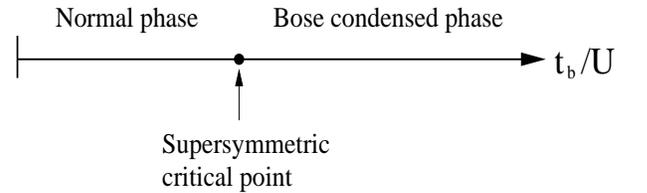}
\caption{
The schematic phase diagram as a function of the ratio of the boson hopping $t_b$ to the on-site 
boson repulsion energy $U$ for a generic value of $h_0$.
}
\label{fig:5}
\end{figure}

The schematic phase diagram of the Hamiltonian (\ref{model}) for a generic value of $h_0$ is shown in Fig. \ref{fig:5} in the parameter space of $t_b/U$.
There is a second order phase transition between the normal phase for small $t_b/U$
and the bose condensed phase for large $t_b/U$.
In the normal phase, the fermions are gapless while a gap opens in the bose condensed phase.
The critical point is described by the supersymmetric Wess-Zumino theory although 
both the normal and the bose condensed phases are non-supersymmetric.

Although the evidences for the emergent supersymmetry, that is, the supersymmetric relation between couplings and the scaling dimensions, are obtained based on the calculation to leading order in $\epsilon$, we expect that the same conclusion holds to all orders in $\epsilon$, at least for small $\epsilon$ for the following reason.
The present model contains the supersymmetric Wess-Zumino theory in the sense that the supersymmetric Wess-Zumino theory can be reached at least by fine tunings of the bare parameters.
Therefore, the Wess-Zumino theory should appear as a fixed point of this model in any case in the parameter region of $\lambda_1 \sim h^2 \sim \epsilon$ although we don't know {\it a priori} whether it is a stable or unstable fixed point.
Since the fixed point which was identified as the Wess-Zumino theory at leading order in $\epsilon$ is the only fixed point which exists in that parameter range, and the higher-order terms in the $\epsilon$-expansion can not generate a new fixed point near the generic stable fixed point, the fixed point obtained to leading order in $\epsilon$ should correspond to the Wess-Zumino fixed point to all orders in $\epsilon$.
In other words, if the fixed point obtained to leading order in $\epsilon$ were not the  supersymmetric Wess-Zumino fixed point, another fixed point which corresponds to the Wess-Zumino theory should have appeared to leading order in $\epsilon$.

Note that the supersymmetry transformations in Eq. (\ref{susy}) 
mix bosons and fermions which carry different global U(1) charges,
where the U(1) charges of the fermions and bosons 
are given by $Q_\psi= 1$ and $Q_\phi = 2$ respectively.
Therefore the supercharges should carry the U(1) charge which
is called R-charge. 
At the supersymmetric critical point, the super-Poincare symmetry is enlarged
to an even bigger symmetry, that is, the superconformal symmetry 
which includes additional fermionic
generators and the R-charge\cite{MINWALLA}.
The additional fermionic generators arise from the commutator
of the supercharges and the conformal generators.
The R-charge which enters in the superconformal algebra 
is related to the global U(1) charge as
$R = - \frac{Q}{3}$.
The factor of $3$ in the definition of the R-charge is due to the
cubic superpotential of the Wess-Zumino theory\cite{ARGYRES}.
Due to unitarity, there exists a constraint on the R-charge and scaling dimension
of an operator.
In 2+1D, the  superconformal symmetry puts a lower bound on 
a scaling dimension as
$D_{{\cal O}} \geq |R_{{\cal O}}|$, 
where $D_{{\cal O}}$ and $R_{{\cal O}}$ 
are the scaling dimension and the R-charge of an operator ${\cal O}$ respectively.
The equality is saturated for a chiral primary field and
the one-loop calculation gives 
the exact anomalous dimensions for the fundamental fields
with $\eta_\phi = \eta_\psi = 1/3$\cite{AHARONY}.
Note that these coincide with the values obtained by putting $\epsilon=1$ in 
the one-loop results of Eq. (\ref{SD}).
Non-chiral primary fields generally receive radiative corrections
and the critical exponent for the order parameter 
is calculated to be
$\nu_\phi = \frac{1}{2} + \frac{\epsilon}{4} + {\cal O}(\epsilon^2)$ 
at leading order in $\epsilon$\cite{SCOTT}.

\section{Conclusion}
In conclusion, we found a 2+1D non-supersymmetric lattice model whose quantum critical point is described by the supersymmetric Wess-Zumino theory in an $\epsilon$-expansion. 
The supersymmetric critical point describes the generic second order phase transition between a normal phase and a bose condensed phase of the bose-fermi mixed system.
The exact anomalous scaling dimensions are predicted.
In principle, the boson can dynamically arise as a Cooper pair in a system which has only fermions as microscopic degrees of freedom.
In such a case, the critical point describes a superconducting phase transition.
Therefore, supersymmetry can emerge at a critical point of a pure fermionic system.
It is of interest to examine such possibility in the future.
Finally, although the supersymmetric theory describes 
the generic second order phase transition in $4-\epsilon$ dimension for a small $\epsilon$,
in 2+1D we can not exclude other possibilities.
For example, the supersymmetric critical point may arise only as 
a multi-critical point due to an occurrence of
other supersymmetry-breaking relevant perturbation, or 
the critical point itself may disappear due to a first order phase transition.

\section{Acknowledgement}
The author thanks M. P. A. Fisher, S. Sachdev, X.-G. Wen and C. Wu for helpful discussions, 
M. Hermele for pointing out the O(4) fixed point, and J. Alicea for proof reading the manuscript and making helpful suggestions.
This research was supported by the National Science Foundation under Grant No. PHY05-51164.






\begin{thebibliography}{27}
\bibitem{C2} In the literature, it is conventional to write the interaction as 
$ \phi_n \psi_n^T \epsilon \psi_n$ by rewriting $\phi \rightarrow \phi^*$ in Eq. (\ref{lfb}). 
Here we keep our notation to remind us that 
a boson is made (`created') out of two fermions in the microscopic model.
\end{thebibliography}
\end{document}